\documentclass[a4paper, 11pt]{article}
\usepackage{graphicx}

\begin{document}

\title{GPS Fit Method for Paths of Non Drunken Sailors and its Connection to Entropy}
\author{Fetze Pijlman\\ \begin{small}fetze.pijlman@signify.com\end{small} \\ \small Signify Research\\ \small High Tech Campus 7, Eindhoven, The Netherlands}

\maketitle

\begin{abstract}
Estimating the altimeters a cyclist has climbed from noisy GPS data is a challenging problem. In this article
a method is proposed that assumes that a person locally takes the shortest path. This results in an algorithm
that does not need smoothing parameters. Moreover, it turns out that this assumption allows one to find a similarity
between entropy and likelihood which results to the introduction of an entropic force.
\end{abstract}

\section{Introduction}

Using GPS data points for fitting paths of cyclists remains at present a challenging problem. Apps
such as strava~\cite{strava} use GPS coordinates to estimate how many meters a cyclist has cycled upwards but it is not uncommon that this estimate can be off by more than 50\% (see also Ref.~\cite{dcrainmaker}). The main reason for this inaccuracy is the uncertainty in the coordinate measurement (where the uncertainty along the different axes x, y, z may be different). Although there are several fit methods available, such as spline fitting, it is the aim of this article to present a new method that has an interesting relation with 
entropy.

\begin{figure}
\begin{center}
\begin{tabular}{ccc}
\includegraphics[width=5cm,angle=0]{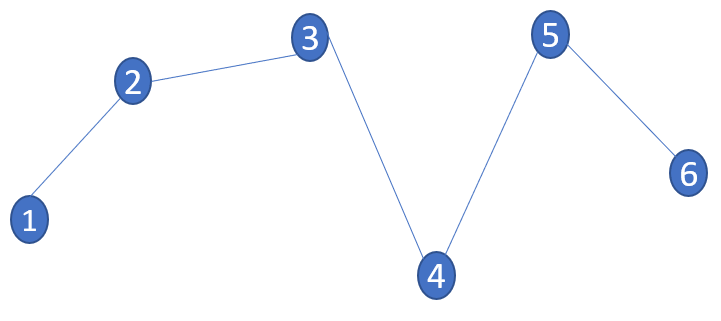}
&\ \ \ &
\includegraphics[width=5cm,angle=0]{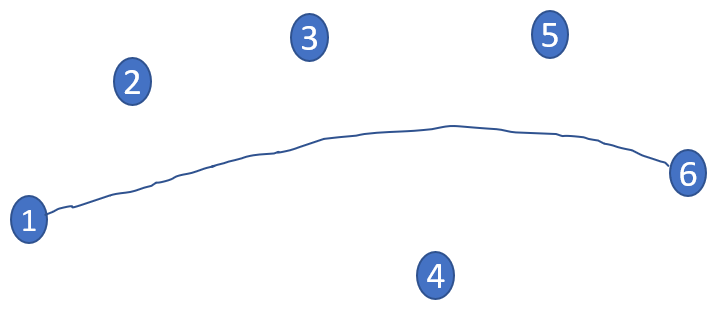}
\end{tabular}
\caption{Illustration of different path fits. In the left illustration the paths are running directly through the
measured points while in the right illustration the paths are running at some distance from the points.\label{illustrationProblem}}
\end{center}
\end{figure}

We can visualize the problem as presented in Fig.~\ref{illustrationProblem}. Somehow one needs to
find the "optimal" path from a set of measurement points. As each point has an uncertainty, it is too 
conservative to draw the path directly through the points. A more balanced path can be achieved by taking this 
uncertainty into account but the question is how to find this balanced path and its justification. The assumption
that is used here is to assume that a person locally takes the shortest path, hence the title of this paper non
drunken sailor (drunken sailor is used for describing a person performing a random walk (see also Ref.~\cite{randomwalk}) whereas here the person takes locally the shortest path). 
Please note that this assumption is sometimes not met, e.g., when strolling around with a family through a shopping center. So the applicability of this paper is limited to the scope of this assumption. The other assumptions, which
normally do apply, are that the uncertainty of measurement points can be described through some probability function and that the number of points is large.

It turns out that the mentioned assumptions above can be translated into a thermodynamic model
in which the fitting process can be described by a statistical force that has its origin from entropy.
Entropic forces are of interest in other fields such as Ref.~\cite{hooft} and Ref.~\cite{verlinde}. An explicit derivation
on how entropic forces can arise in fitting methods is insightful when studying other fields as well.

\section{Problem description}

We will assume that the uncertainty between the measured position $c_i$ and the true position $r_i$
can be described by some probability function $P(r_i)$. 
The expected likelihood can be computed using the definition for expectation
\begin{equation}
\left< P \right> = \int P(r_i) P(r_i) {\rm d}r_i,\label{expected_likelihood}
\end{equation}
where $r_i$ is the coordinate and which may have several dimensions.
In the following we will assume that the expectation value for likelihood $<P>$ is the same for all data points (although one can actually allow for point dependent expectation values for probability).
The likelihood of the data can be computed by multiplying the likelihood of all data points. In the limit of
$n \to \infty$ this likelihood equals the expected likelihood
\begin{equation}
\lim_{n\to \infty}\prod_{i=1}^n \ P(r_i) = \left< P \right>^n.\label{constraint}
\end{equation}
The equation above will serve as a constraint in the fitting process.

\section{Fit method}

The minimization of the path length can be achieved by imagining a string that runs through some points that we will
call control points where the control points are connected to the measurement points via springs. 
When a string is completely loose the springs are completely contracted which makes that the string runs through 
the measurement points. However, when starting to pull the string (effectively making it shorter), the springs will start to execute a force which becomes higher the further the control point is away from the measurent point. The optimal path is achieved at the moment when the constraint in Eq.~\ref{constraint} is
met.

We observe that the optimal path has the highest likelihood for that particular string length, otherwise an even shorter path was feasible. This justifies the postulation of an entropic force
that forces that path to the highest likelihood at fixed path lengths. This postulate has similarities with entropy which
for an isolated system can only stay equal or increase.

Let's study what happens when we pull the string away from a certain point j. From now on we will assume that one has carried out a linear coordinate transformation for each $r_j \rightarrow r_j + c_j$ such that all $c_j$'s are dropped. 
When moving the string over an infinitesimal distance $\Delta r_j$ one changes the length of the overal string by 
\begin{equation}
\Delta L = \frac{{\rm d}L}{{\rm d} r_j} \Delta r_j.\label{changeLength}
\end{equation}
Throughout this article the coordinate systems are chosen such that for a positive $\Delta r_j$ the length $L$
is shorted so that $\Delta L < 0$.
By moving the string over an infinitesimal distance one also changes the likelihood which is most conviniently computed by considering the log of the likelihood
(without loosing generality), giving
\begin{equation}
\Delta \log \left(  \prod_{i=1}^n \ P(r_i)  \right) = \frac{{\rm d}\log P(r_j)}{{\rm d} r_j}\ \Delta r_j.\label{changeLogLikelihood}
\end{equation}

Now that we know how the overal log likelihood changes when we pull the string from one point, we can use the entropic force postulate to compute the new situation. Please note that the postulate of highest likelihood means also highest log 
likelihood. At fixed length $L$ one can compensate the change $\Delta L$ from one point by a change $-\Delta L$ from another point. By the postulate, such changes will occur if it would increase the overal likelihood. In effect, all such changes will occur until the maximum
likelihood is reached which is the case when the change in log likelihood per $\Delta L$ is the same for all points. In other words, in the optimum state the forces are in balance:
\begin{equation}
\forall j: \frac{\Delta \log \left(  \prod_{i=1}^n \ P(r_i)  \right)}{\Delta L} = F_S(L) =  \frac{\frac{{\rm d}\log P(r_j)}{{\rm d} r_j}}{\frac{{\rm d}L}{{\rm d} r_j}},
\end{equation}
where $F_S(L)$ is some constant. When one rewrite the expression above one obtains
\begin{equation}
F_S(L) \frac{{\rm d}L}{{\rm d} r_j} = \frac{{\rm d}\log P(r_j)}{{\rm d} r_j}.\label{forceEquation}
\end{equation}
When we look at the equation above we recognize the left-hand-side as the resulting force from the string tension $F_S(L)$
which is balanced by the entropic force which is apparently the derivative of the log probability distribution. The algorithm for finding the optimal
path now becomes rather simple. 1) Start with a loose string, 2) create a small pull in the string by pulling it off from the first point, 3) let the system readjust itself via the entropy forces, 4) compute whether constraint has been met,
if yes then the optimal path has been found, if not continue with step 2.

\section{Example: entropic force in two dimension with for Lorentz distributed uncertainties}

The probability distribution of the measurement error may take several different forms. In this section we will consider
the Lorentz (or Cauchy) distribution. The distribution reads
\begin{equation}
P_i(\vec{x}_i) = \frac{1}{\pi \gamma} \left( \frac{\gamma^2}{\left| \vec{x}_i - \vec{c}_i \right| + \gamma^2} \right).
\end{equation}
For the expected likelihood (see Eq.~\ref{expected_likelihood}) one finds
\begin{equation}
\left< P \right> = \frac{1}{\pi},
\end{equation}
which for $n$ points becomes
\begin{equation}
\left< \prod_i \frac{1}{\pi \gamma} \left( \frac{\gamma^2}{\left| \vec{x}_i - \vec{c}_i \right| + \gamma^2} \right) \right>
= \left( \frac{1}{\pi}  \right)^n.
\end{equation}
For $n$ going to infinity one simply gets
\begin{equation}
\lim_{n \to \infty} \prod_i \frac{1}{\pi \gamma} \left( \frac{\gamma^2}{\left| \vec{x}_i - \vec{c}_i \right| + \gamma^2} \right) =\lim_{n \to \infty}  \left( \frac{1}{\pi}  \right)^n.
\end{equation}
Please note that the relation above is expected to be met for large $n$ if the assumption of having a Lorentz distribution
applies.

\begin{figure}
\begin{center}
\includegraphics[width=4cm,angle=0]{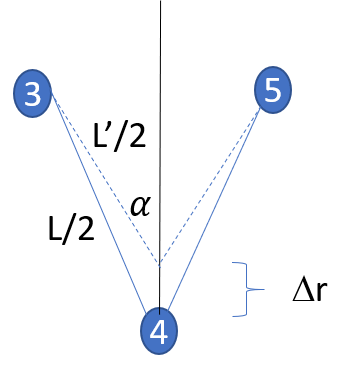}
\caption{Shortening the path by pulling path away from point 4.\label{force}} 
\end{center}
\end{figure}

Let's study what happens when we pull the string away from a certain point such that the overal length is shortened. 
In Fig.~\ref{force}
one can read off the length reduction giving
\begin{equation}
\Delta L = \frac{{\rm d}L}{{\rm d} r} \Delta r =   -2 \Delta r\ \cos \alpha.
\end{equation}
A reduction of the overal length is accompanied by a reduction in the log likelihood which is
\begin{equation}
\Delta \log \left(  \prod_{i=1}^n \ P(\vec{x}_i)  \right) = \frac{{\rm d}\log P(\vec{x}_j)}{{\rm d} x_j}\ \Delta x_j
= -\ \frac{2r}{\gamma^2+r^2} \Delta r.
\end{equation}

\begin{figure}
\begin{center}
\includegraphics[width=4cm,angle=0]{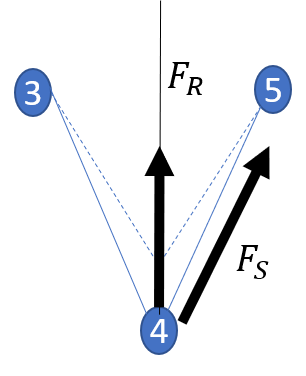}
\caption{Shortening the path by pulling path away from point 4.\label{string_tension} }
\end{center}
\end{figure}

As discussed in the previous section leading to Eq.~\ref{forceEquation}, at a fixed overal length the maximum likelihood path is found where all the 
forces are in balance. This means that for all points the ratio of the change in log likelihood with change in L is constant
\begin{equation}
\forall j: 2\cos \alpha_j F_S(L) = \frac{2r}{\gamma^2+r^2}\label{force2d},
\end{equation}
where $F_S(L)$ is some constant.

When comparing Eq.~\ref{force2d} with Fig.~\ref{string_tension} the analogy with mechanical forces becomes quite strong.
On the left-hand-side of Eq.~\ref{force2d} we see the net force $F_R$ as a result of the string tension which is balanced
by the postulated entropic force which for a Lorentz distribution takes the form of the right-hand-side of Eq.~\ref{force2d}. The concept of string tension originates for mechanics, see e.g. Ref.~\cite{mechanics}, and appears here in a different context.

\section{Example: entropic force in one dimension with for normal distributed uncertainties}

In this section we study the problem that was discussed in the introduction. How to compute the amount of altimeters when assuming that a person chose the path with the least amount of height differences.
As an example we will consider height measurements $c_i$ for we will assume that the error is normally distributed.  The 
probability distribution in this case reads
\begin{equation}
P_i(z_i) = \frac{1}{\sqrt{2\pi} \sigma}\ e^{\frac{-(z_i - c_i)^2}{2\sigma^2}},
\end{equation}
where $\sigma$ is the standard deviation. The expected likelihood can be computed using Eq.~\ref{expected_likelihood} giving
\begin{equation}
\left< P \right> = \frac{1}{2 \sigma \sqrt{\pi}}.
\end{equation}
So for $n$ points Eq.~\ref{constraint} now becomes
\begin{equation}
\lim_{n\to \infty} \prod_{i=1}^n \frac{1}{\sqrt{2\pi} \sigma}\ e^{\frac{-(z_i - c_i)^2}{2\sigma^2}} = \lim_{n \to \infty} \left( \frac{1}{2 \sigma \sqrt{\pi}} \right)^n.
\end{equation}

\begin{figure}
\begin{center}
\begin{tabular}{ccc}
\includegraphics[width=4cm,angle=0]{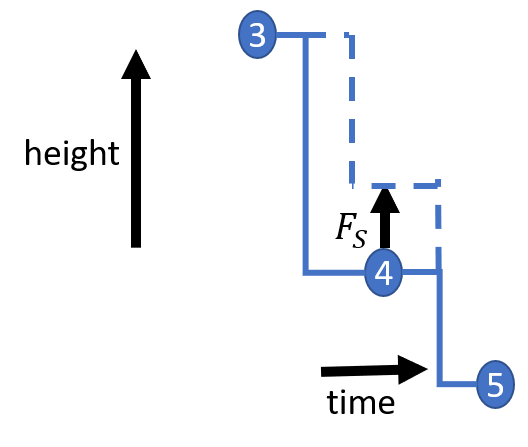}&&
\includegraphics[width=4cm,angle=0]{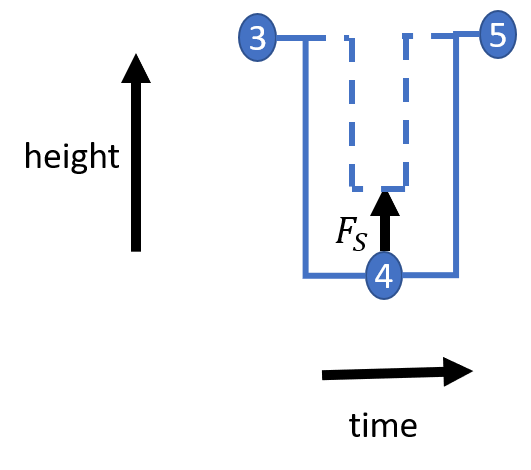}
\end{tabular}
\caption{Two examples when pulling path away from point 4. 
In the left figure point 4 has its direct neighbours on both sides in which case a pull on point 4 does not lead to a shortening of the overal vertical path (class A).
In the right figure, point 4 has its direct neighbours (point 3 and point 5) on the same side. Consequently, a pull from point 4 leads to a shortening of the overal vertical path (class B).
\label{forceHeight}} 
\end{center}
\end{figure}

Similar to the previous section we study what happens when one pulls a control point from a measurement point (please see the illustration in Fig.~\ref{forceHeight}) but in this case there turns out to be a small complication.
Any given path runs through control points points that are close to the measurement points.
It turns out that for a given path one can divide the control points in the following two classes. For the first class,
that we will call class A, a small change in the position of the control point does not lead to a change in the overal
vertical length. For the second class, that we will call class B, a small change in the control point does lead to a
change in the overal vertical length.

Considering now only the points of class B one finds that
Eq.~\ref{changeLength} becomes now ($\Delta L$ is how the change in altimeters which is the sum of changes up and down)
\begin{equation}
\Delta L = \frac{{\rm d}L}{{\rm d} r_j} \Delta r_j = - 2 \Delta r_j.
\end{equation}
The change in the log likelihood (see also Eq.~\ref{changeLogLikelihood}) becomes for class B simply
\begin{equation}
\Delta \log \left(  \prod_{i=1}^n \ P(r_i)  \right) = \frac{{\rm d}\log P(r_j)}{{\rm d} r_j}\ \Delta r_j = - \frac{\left( z_i - c_i \right)}{\sigma^2},
\end{equation}
after which the force equation (Eq.~\ref{forceEquation}) becomes
\begin{equation}
\forall i:\ 2 F_S(L) = \frac{\left( z_i - c_i \right)}{\sigma^2}.
\end{equation}
The resulting force from the maximum likelihood postulate is similar to a linear spring. Points that are further away from the measurement point experience a larger force than points that are closer to their measurement point. In the 
equilibrium state, all these points should experience the same force.

The algorithm for computing the minimal vertical path can now be summarized as follows:
\begin{enumerate}
\item initiate path by putting all control points at the measurement points
\item shorten the overal path length by pulling the control points of class B by a small amount away from the measurement points
\item compute for each control point of class B the force (which is linear to the distance of the path and measurement point)
\item if all forces of previous step are equal, then continue with step 6.
\item move the control point that experiences the highest force towards the measurement point by a fixed amount and move by the same amount the control point that experiences the lowest force away from the measurement point and continue with step 3.
\item compute the overal likelihood of the $n$ points. If equal or lower than the expectation likelihood then one has
obtained the optimal path, else go to step 2.
\end{enumerate}
Please note that the number control points $n$ of type class B will fluctuate during the execution of the algorithm.

\section{Conclusions}

There are various methods available for estimating on the basis of noisy
GPS data how many meters a cyclist has climbed during a ride but apparently it is not an easy task to carry out.
In this article a new method is proposed for persons that locally take the shortest path. The assumption allowed
us to construct a theoretical entropic force which enables us to find the actual/expected path a person has taken.

For the two dimensional example the shape of the path is defined by the control points which can be interpreted as
being kept close to the measurement point by a spring. By increasing the string tension (shortening of the path), the 
springs are elongated until one meets the expectation value. The obtained path meets the expectation value which
should be close to reality as the number of measurement points for GPS tracks is often high (GPS points are often
measured every second). The fact that the obtained path has strong bents (which seem unnatural) is a short coming of the model as the model does not include an upper limit on acceleration. This aspect can be included in future studies.

The one dimensional example which is relevant to the motivation of this article turned out to be slightly more complicated
as the points need to be separated in classes. However, the same methodoly could also be used here leading to a nice
short algorithm that can hopefully contribute in a better estimation of number altimeters climbed.

Besides the concrete problem of finding a path given GPS data points, the introduction of an entropic force that finds its
origin in statistics is of interest. Just like entropy can only increase for an isolated system, one can postulate
that at fixed path lengths the likelihood will only increase (or stay equal).

\end{document}